\documentstyle[sprocl]{article}

\def\be{\begin{equation}}
\def\ee{\end{equation}}
\def\bea{\begin{eqnarray}}
\def\eea{\end{eqnarray}}
\begin{document}
\def\lambdabar{{\mathchar'26\mkern-9mu\lambda}}

%To Prof Nick Karayiannis -- do read this:-
%If needed the word of Chapter~1, you can type in at the 
%\title{}. The words will be in caps and lowercase. 
%For chapter title can be in all caps or in caps and lowercase.
%It is up to the author to type for the case sensitive but 
%all articles must be in the same style. 
%But mostly for Review Volume are without this Chapter~1.
%Thank you
%Jessie   13/4/2000

\title{GRAVITATION AND NONLOCALITY}

\author{BAHRAM MASHHOON}

\address{Department of Physics and Astronomy,\\ University of Missouri-Columbia,\\
Columbia, Missouri  65211, USA\\E-mail: mashhoonb@missouri.edu} 


\maketitle\abstracts{ The physical basis of the standard theory of general
relativity is examined and a nonlocal theory of accelerated observers is described
that involves a natural generalization of the hypothesis of locality.  The nonlocal
theory is confronted with experiment via an indirect approach.  The implications of
the results for gravitation are briefly discussed.}

\section{Introduction} Einstein's general theory of relativity [1] is a successful
classical theory of gravitation inasmuch as it agrees with all experimental data
available at present [2].  To develop a microphysical gravitation theory, it may
prove useful to investigate the extent to which the basic physical tenets of
general relativity conform with the fundamental principles of the quantum theory. 
This is the general approach adopted in this work.  

The basic physical assumptions underlying general relativity can be determined by
investigating the measurement problem in the theory of relativity.  In this way one
can characterize the observational assumptions that are necessary in order to
provide a physically consistent interpretation of all relativistic formulas [3]. 
The resulting four basic pillars of the theory are the following:

(i) Lorentz Invariance,\\

(ii) Hypothesis of Locality,\\

(iii) Einstein's Principle of Equivalence,\\

(iv) Correspondence with Newtonian Gravitation and the Gravitational Field
      Equations.

Briefly, the inhomogeneous Lorentz transformations in (i) relate the measurements
of physical quantities by ideal inertial observers in Minkowski spacetime. 
However, observers in Minkowski spacetime are generally accelerated, at least along
a portion of their worldlines.  To deal with realistic (i.e. accelerated)
observers, the standard theory of relativity contains the assumption that an
accelerated observer is at each instant momentarily equivalent to a hypothetical
comoving inertial observer; that is, the acceleration of the observer is considered
{\it locally} immaterial (``Hypothesis of Locality'').  Moreover, an observer in a
gravitational field is presumed to be {\it locally} equivalent to a certain
accelerated observer in Minkowski spacetime according to Einstein's heuristic
principle of equivalence.  Assumptions (ii) and (iii) taken together imply that
observers in a gravitational field are all locally inertial.  The simplest way to
connect these local inertial frames is via Riemannian geometry of curved spacetime,
where the curvature is identified with the gravitational field and free test
particles and null rays follow geodesics of the spacetime manifold.  Finally, in
the geometric framework of general relativity, Einstein's field equations are the
simplest equations that provide a natural and consistent generalization of
Poisson's equation of Newtonian gravity.  

Dirac's generalization of Schr\"{o}dinger's equation and the subsequent
achievements of relativistic quantum field theory indicate that (i) can be
integrated into the general framework of quantum theory.  Therefore, the main part
of this paper is concerned with the status of (ii) vis-a-vis quantum mechanics. 
Section 2 examines the physical basis of (ii) and section 3 describes a nonlocal
theory of accelerated observers.  The observational aspects of this nonlocal theory
are examined in section 4.  In conclusion, some of the implications of these ideas
for the theory of gravitation are briefly described in section 5.  

\section{Hypothesis of Locality} The ultimate physical basis of the hypothesis of
locality is Newtonian mechanics, where the {\it state} of a particle is determined
by its position and velocity.  Thus the accelerated observer and the otherwise
identical hypothetical instantaneously comoving inertial observer are equivalent,
since they both share the same state.  

In terms of realistic measurements of the accelerated observer, the hypothesis of
locality would hold if these measurements are essentially pointwise and
instantaneous, so that the influence of inertial effects can be neglected over
length and time scales that are characteristic of elementary local measurements. 
To state this criterion in a quantitative way, let us first note that an
accelerated observer can be characterized by certain acceleration lengths ${\cal
L}$ that involve the speed of light $c$ and certain scalars formed from the
translational and rotational accelerations of the observer.  If $\lambdabar$ is the
intrinsic length scale of the phenomenon under observation, then $\lambdabar/{\cal
L}$ characterizes the expected deviation from the hypothesis of locality.  For
instance, in a laboratory fixed on the rotating Earth, the typical acceleration
lengths would be $c^2/g_{\oplus} \simeq 1$ lyr and $c/\Omega_{\oplus}\simeq 28
\,{\rm AU}$; therefore, for most experimental situations $\lambdabar/{\cal L}$
would be negligibly small.  It follows that the hypothesis of locality is
approximately valid under most current observational situations.  Measuring
devices that, like the rods and clocks of classical relativity theory (cf. [1], p.
60), obey the hypothesis of locality are called ``standard''; hence, a standard
clock measures proper time along its worldline. 

To delve deeper into the various limitations of the hypothesis of locality, we
consider a thought experiment involving the reception of a normally incident plane
electromagnetic wave by a rotating observer.  Let us choose a global inertial
system in which the observer moves with constant frequency $\Omega$ in the $(x,
y)$-plane on a circle of radius $R$ about the origin such that $x=R\:{\rm
cos}\:\varphi$ and $y=R\:{\rm sin}\:\varphi$, where $\varphi=\Omega t$.  Moreover,
the monochromatic plane wave propagates along the $z$-axis with frequency
$\omega$.  

It follows from the hypothesis of locality that the natural orthonormal tetrad
frame of the rotating observer can be given with respect to the inertial frame by
[3]

\begin{equation}
\lambda^{\mu}_{{\:\:}(0)} =\gamma(1, -\beta\;{\rm sin}\;\varphi,\; \beta\;{\rm
cos}\;\varphi,
\;0)\;\;,
\end{equation}

\begin{equation}
\lambda^{\mu}_{{\:\:}(1)} = (0,\;{\rm cos}\;\varphi\;,\;{\rm
sin}\;\varphi\;,0)\;\;,
\end{equation}

\begin{equation}
\lambda^{\mu}_{{\:\:}(2)} = \gamma(\beta,\;-{\rm sin}\;\varphi\;,\;{\rm
cos}\;\varphi,\;0)\;\;,
\end{equation}

\begin{equation}
\lambda^{\mu}_{{\:\:}(3)} = (0\;,\;0\;,\;0\;,\;1)\;\;,
\end{equation}

\noindent where $\gamma$ is the Lorentz factor, $\gamma = (1-\beta^2)^{-1/2}$, and
$\beta =v/c=R\Omega/c$.  The motion of the frame along the worldline can in general
be characterized by six scalar quantities that form an antisymmetric tensor
$\phi_{\alpha\beta}$ defined by
$d\lambda^{\mu}_{{\,\,}(\alpha)}/d\tau=\phi_{\alpha}^{{\;\;}\beta}\lambda^{\mu}_{{\;}(\beta)}$. 
Here
$\tau$ is the proper time, $d\tau = \gamma^{-1}\;dt$, the ``electric'' part of
$\phi_{\alpha\beta}$ corresponds to translational acceleration ${\tilde{g}}_i =
\phi_{0i}$ and the ``magnetic'' part $\tilde{\Omega}_i = {1\over
2}\epsilon_{ijk}\phi^{jk}$ corresponds to the rotational frequency of the spatial
frame.  For the tetrad (1) - (4), we find that the scalars $\tilde{g}_1 = -\gamma^2
R\Omega^2$ and $\tilde{\Omega}_3=\gamma^2\Omega$ are the only nonzero components of
${\bf\tilde{g}}$ and ${\bf\tilde{\Omega}}$ corresponding respectively to a
centripetal acceleration of magnitude $\gamma^2v^2/R$ and a rotation of the spatial
frame about the $z$-axis of frequency $\gamma^2\Omega$.  It should be noted that
{\it proper} acceleration scales can be constructed from the invariants
${1\over 2}\phi_{\alpha\beta}\phi^{\alpha\beta}$ and ${1\over
2}\phi^{*}_{\alpha\beta}\phi^{\alpha\beta}$, where $\phi^{*}_{\alpha\beta}$ is the
dual of $\phi_{\alpha\beta}$; in the case of uniform rotation, these are
respectively $\gamma^2\Omega^2$ and zero.  It follows that ${\cal
L}=c/(\gamma\Omega)$, where $\gamma\Omega=d\varphi/d\tau$ is the proper rotation
frequency of the observer [3].  

Regarding the reception of the wave by the observer, we note that an
electromagnetic field may be represented in terms of the components of the Faraday
tensor
$f_{\mu\nu}$ as measured by the standard set of ideal inertial observers at rest in
the underlying global frame.  The field measured by an arbitrary accelerated
observer is then the projection of $f_{\mu\nu}$ on the orthonormal tetrad of the
observer 

\begin{equation}
\hat{f}_{\alpha\beta} =
f_{\mu\nu}\;\lambda^{\mu}_{{\;}(\alpha)}\;\lambda^{\nu}_{{\;}(\beta)}\;\;.
\end{equation}

\noindent It is possible to express (5) using the six-vector representation of the
Faraday tensor, i.e. in terms of the electric and magnetic fields ---
$f_{\alpha\beta}\rightarrow({\bf E}, {\bf B})$, as

\begin{equation} {\hat f}=\Lambda f\;\;.
\end{equation}

\noindent The incident electromagnetic radiation field under consideration may
therefore be expressed as the real part of 

\begin{equation} f=i\omega A \left[ \begin{array}{c} {\bf e}_\pm \\ {\bf b}_\pm
\end{array}\right]\;\;e^{-i\omega(t-z/c)}\;\;,
\end{equation}

\noindent where $A$ is a complex amplitude, ${\bf e}_\pm = ({\bf e}_1 \pm i{\bf
e}_2)/\sqrt{2}, {\bf b}_\pm = \mp i{\bf e}_\pm$ and the upper (lower) sign
indicates radiation of positive (negative) helicity.  Here ${\bf e}_1$ and
${\bf e}_2$ are unit vectors along the positive $x$ and $y$ directions,
respectively.  Using equations (1) - (7), we find that the field measured by the
rotating observer is given by the real part of 

\begin{equation} {\hat f} = i\gamma\omega A\left[ \begin{array}{c} {\bf\hat
e}_{\pm} \\ {\bf\hat b}_{\pm}\end{array}\right]\;\;e^{-i{\hat\omega}\tau}\;\;,
\end{equation}

\noindent where ${\bf\hat b}_\pm = \mp i{\bf\hat e}_\pm$ and

\begin{equation} {\bf\hat e}_\pm = \frac{1}{\sqrt{2}}\; \left[ \begin{array}{cc} 1
\\ \pm i\gamma^{-1}
\\ \pm i\beta
\end{array}\right]
\end{equation}

\noindent involve unit vectors with respect to the tetrad axes, $\gamma\tau=t$ and 

\begin{equation} {\hat\omega} = \gamma(\omega\mp\Omega)\;\;.
\end{equation}

To interpret these results, let us first note that a simple application of the
hypothesis of locality would connect the instantaneous inertial frame of the
rotating observer with the global inertial frame.  In terms of the propagation
four-vector of the wave, the result is the transverse Doppler effect; namely, the
accelerated observer measures a frequency $\omega^{\prime}=\gamma\omega$, where the
Lorentz factor accounts for time dilation.  However, if the hypothesis of locality
is applied to the field and the result is Fourier analyzed --- a nonlocal operation
in proper time --- then, the result is equation (10), which goes beyond
$\gamma\omega$ by terms of the form $\Omega/\omega = \lambdabar^{\prime}/{\cal
L}$.  Therefore, ${\hat\omega}\rightarrow\omega^{\prime}$ in the JWKB limit
$\Omega/\omega\rightarrow 0$.

Equations (8) - (10) have a simple intuitive interpretation : For positive
(negative) helicity radiation, the electric and magnetic fields rotate with
frequency $\omega\;(-\omega)$ about the direction of propagation of the wave and
the rotating observer perceives positive (negative) helicity radiation but with
frequency $\omega - \Omega\;(\omega + \Omega)$ augmented with the time dilation
factor $\gamma$.  Partial experimental evidence is presented in [4] for this
phenomenon; it is an example of the general spin-rotation coupling (see [5 - 9] for
discussions and reviews).  On the other hand, equation (10) has a remarkable
consequence for which there is no observational evidence:  for incident positive
helicity radiation of frequency $\omega = \Omega$, the radiation field becomes
static, i.e. the wave stands completely still, for the whole class of observers
rotating uniformly with frequency $\Omega$ about the $z$-axis.  More generally, for
an obliquely incident radiation field ${\hat\omega} = \gamma(\omega - M\Omega)$,
where $M$ is the multipole parameter associated with the $z$-component of the {\it
total} angular momentum of the field; therefore, the radiation can stand completely
still with respect to the rotating observers for $\omega = M\Omega$.  

Another general consequence of the hypothesis of locality reflected in equations
(5) - (10) is the following:  If the incident radiation is a linear superposition
of the two possible helicity states, then the radiation as seen by the rotating
observer has the {\it same} amplitudes in terms of the transformed basis (9) but
different frequencies.  It would also be interesting to confront this prediction of
the hypothesis of locality regarding {\it helicity amplitudes} with observation. 

\section{Nonlocal Theory of Accelerated Observers}  Consider anew the reception of
electromagnetic radiation by an accelerated observer.  How is the class of
momentarily comoving inertial observers that measure ${\hat
f}_{\alpha\beta}(\tau)$ related to the accelerated observer that measures
$F_{\alpha\beta}(\tau)$ while passing through the continuous infinity of local
inertial systems?  The hypothesis of locality postulates that $F_{\alpha\beta}$ is
equal to ${\hat f}_{\alpha\beta}$ at each instant of proper time $\tau$.  However,
the most general linear and causal relationship between ${\hat
f}_{\alpha\beta}(\tau)$ and the field
$F_{\alpha\beta}(\tau)$ that is actually measured by the accelerated observer is
given by a Volterra integral equation

\begin{equation} F_{\alpha\beta}(\tau)={\hat f}_{\alpha\beta}(\tau) +
\int^{\tau}_{\tau_0}\;K_{\alpha\beta}\;^{\gamma\delta} (\tau,
\tau^{\prime}){\hat f}_{\gamma\delta} (\tau^{\prime})d\tau^{\prime}\;\;,
\end{equation}

\noindent where $\tau_0$ is the instant at which the acceleration is turned on. 
It follows from Volterra's theorem that in the space of continuous functions the
relationship between $f=\Lambda^{-1}{\hat f}$ and $F$ given by equation (11) is
unique; this uniqueness result has been extended to square-integrable functions by
Tricomi [10].  Equation (11) is manifestly Lorentz covariant as it deals only with
scalar quantities.  It remains to determine the kernel
$K_{\alpha\beta\gamma\delta}$ in terms of the acceleration of the observer.  

Let us note that if the kernel $K$ is simply proportional to the acceleration, then
the magnitude of the nonlocal part in equation (11) would generally be of the form
$\lambdabar^{\prime}/{\cal L}$, as expected.  The basic approach to the
determination of the kernel in this theory is to exclude the possibility that an
incident radiation field could stand completely still with respect to an
accelerated observer (cf. section 2).  In the case of ideal inertial observers,
this comes about because no observer can move at the speed of light; therefore, it
follows from the Doppler formula
$\omega^{\prime}=\gamma\omega(1-\mbox{\boldmath${\beta}$}\cdot{\bf n})$ for
radiation propagating along a unit vector
${\bf n}$ that if $\omega^{\prime}=0$, then $\omega = 0$.  We demand a similar
outcome for all observers; that is, if in equation (11) $F_{\alpha\beta}$ is
constant, then the incident field $f_{\alpha\beta}$ must be constant.  Imposing
this requirement, equation (11) in six-vector notation reduces to 

\begin{equation}
\Lambda_0 = \Lambda(\tau)+\:\int^{\tau}_{\tau_0}\;K(\tau,
\tau^{\prime})\Lambda(\tau^{\prime})d\tau^{\prime}\;\;,
\end{equation}

\noindent where $\Lambda_0 = \Lambda(\tau_0)$ is a constant $6\times 6$ matrix. 
The Volterra-Tricomi uniqueness theorem now ensures that true incident radiation
fields will never be found to stand completely still by any observer [11].  

Equation (12) is not sufficient to determine a unique kernel $K$; other simplifying
assumptions are necessary.  To this end, let us suppose that $K(\tau,
\tau^{\prime})$ is only a function of one variable.  Two possible situations [12 -
13] are of interest $K(\tau, \tau^{\prime}) = k(\tau^{\prime})$ or
$\tilde{k}(\tau-\tau^{\prime})$.  A detailed analysis reveals that of these two
possibilities only the former (``kinetic'') kernel is acceptable, since the latter
convolution kernel can lead to divergences for nonuniform acceleration [14].  It
follows from equation (12) that the kinetic kernel is directly proportional to the
acceleration of the observer and is given by 

\begin{equation} k(\tau) = -\frac{d\Lambda(\tau)}{d\tau}\;\Lambda^{-1}(\tau)\;\;.
\end{equation}

\noindent In this case, equation (11) can be written as

\begin{equation} F = {\hat f} + \int^{\tau}_{\tau_0}\;k(\tau^{\prime}){\hat
f}(\tau^{\prime})\;d\tau^{\prime}\;\;,
\end{equation}

\noindent so that the nonlocal contribution to the field is a weighted average over
the past history of the accelerated observer.  This circumstance is consistent with
the observation of Bohr and Rosenfeld that the electromagnetic field cannot be
measured at one spacetime point; an averaging process is necessary [15].  From this
standpoint, the kinetic kernel $k$ appears to be unique [16].  

The nonlocal theory of accelerated systems is consistent with the observed absence
of an elementary scalar (or pseudoscalar) particle in nature.  For a scalar field
$\Lambda = 1$, hence $k=0$ and the theory is local.  Thus it would in general be
possible for an observer to stay completely at rest with a pure scalar radiation
field, a possibility that is excluded by our physical postulate.  Hence the theory
predicts that any scalar (or pseudoscalar) particle would have to be a composite.

To illustrate the nonlocal theory, we return to the thought experiment discussed in
detail in section 2.  Let us suppose that for $t<0$ the observer moves uniformly in
the $(x,y)$-plane such that $x=R$ and $y=R\Omega t$ and at $t=\tau=0$ begins the
rotational motion discussed before.  Then the kinetic kernel $k$ given by equation
(13) turns out to be a constant for the case of uniform rotation 

\begin{equation} k=\left[ \begin{array}{cc} k_r & k_t \\ -k_t & k_r
\end{array}\right]\;\;,
\end{equation}

\noindent where $k_r$ and $k_t$ are $3\times 3$  matrices given by $k_r={\tilde{\bf
\Omega}}\cdot{\bf I}=\gamma^2\Omega I_3$ and
$k_t = -{\bf\tilde{g}}\cdot{\bf I} = \gamma^2 \beta\Omega I_1$.  Here $I_i\;,\;
(I_i)_{jk}=-\epsilon_{ijk}$, is a matrix proportional to the operator of
infinitesimal rotations about the
$x^i$-axis.  The radiation field according to the rotating observer is given by the
real part of 

\begin{equation} F={\hat f}\;\frac{\omega\mp\Omega
e^{{i{\hat\omega}\tau}}}{\omega\mp\Omega}\;\;,
\end{equation}

\noindent where ${\hat f}$ is given by equation (8).  Two consequences of
nonlocality should be noted here:  For positive helicity radiation with $\omega =
\Omega$, the result (16) has the character of resonance and $F$ turns out to be a
linear function of proper time $\tau$.  We note that this linear growth (and
eventual divergence) of the field with time would be absent for any finite incident
{\it wave packet}.  Moreover, as a direct result of the fact that $k$ is constant
and the nonlocal part of the field in equation (14) involves an integration over
time, the frequencies $\omega\mp\Omega$ appear in the denominator resulting in a
larger (smaller) measured amplitude for the positive (negative) helicity radiation
as in equation (16).  

Let us recall from the results of section 2 that according to the hypothesis of
locality incident waves of frequency $\omega$ with opposite helicities and equal
amplitudes will be measured by the rotating observer to have equal amplitudes but
different frequencies ${\hat\omega}=\gamma(\omega\mp\Omega)$.  Thus for the {\it
relative} amplitudes of the two helicity states, the radiation field is not
affected by the rotation of the observer according to the standard theory of
relativity.  The nonlocal theory predicts, however, that the field strength will be
higher (lower) when the electromagnetic field rotates in the same (opposite) sense
as the rotation of the observer by the factor $1+\Omega/\omega\;(1-\Omega/\omega)$
for $\Omega/\omega << 1$.  For instance, $\Omega/\omega\sim 10^{-7}$ for radio
waves with $\lambda\sim 1$ cm  incident on a system rotating at a rate of 500
rounds per second.  

This helicity dependence of the amplitude of the radiation field is a definite
signature of nonlocality and the next section is devoted to a discussion of this
effect.

\section{Confrontation with Experiment}  The whole observational basis of the
theory of relativity involves experiments performed in accelerated systems of
reference; however, the acceleration scales are typically very large compared to
the intrinsic scales that are relevant in such experiments and hence rather high
levels of observational accuracy would be needed in order to detect nonlocal
phenomena.  In planning such high-sensitivity experiments, a new problem would be
encountered:  one must consider the influence of acceleration on the accelerated
measuring devices as well.  In view of these difficulties, it is interesting to
explore a novel approach based on the correspondence principle suggested by Steven
Chu [17]:  Under appropriate circumstances, the electrons in atoms may be viewed
as ``accelerated observers''; the predictions of the classical theory could then
be compared with quantum mechanics.  

This idea can be developed in connection with the thought experiment concerning the
measurement of the electromagnetic field of a normally incident wave by a uniformly
rotating observer as follows:  Let us imagine the photoionization process involving
the absorption of an incident photon by an electron bound by a potential and the
subsequent ejection of the electron from the system.  To simulate our thought
experiment (cf. section 2) in the quantum domain, we consider the nonrelativistic
motion of the electron on a ``circular orbit'' in the hydrogen atom, so that the
stationary state of the electron is specified by the quantum numbers $n>1, l=n-1$
and $m=l$; initially, the photon is incident on this state along the
$z$-direction. The electron spin is neglected.  It is interesting to note that
the {\it impulse approximation}, originally suggested by Fermi in treating certain
problems in quantum scattering theory [18], is the quantum analogue of the
hypothesis of locality in this case.  Just as in the hypothesis of locality the
accelerated observer is at each moment replaced by an otherwise identical free
inertial observer, the impulse approximation in effect replaces the bound electron
by a free electron of definite momentum [19].  It then follows that the cross
section for photoionization in this approximation is independent of the helicity
of the incident radiation.  However, the helicity dependence enters the
calculation once the Coulomb interaction is properly taken into account in the
final state.  A detailed treatment reveals that the quantum results and the
predictions of the nonlocal theory are in {\it qualitative} agreement [20].  For
instance, for $mc^2>>\hbar\omega>E_n$, where $m$ is the electron mass, $-e$ is its
electric charge and $-E_n=-me^4/(2\hbar^2n^2)$ is the electron energy in its
initial ``circular orbit,'' the total cross sections $\sigma_+$ (positive helicity)
and $\sigma_-$ (negative helicity) for photoionization in the dipole approximation
are such that

\begin{equation}
\frac{\sigma_-}{\sigma_+} = \frac{3(n-1)+2n\eta}{2n\left[
2(n-1)^2+n(2n-1)\eta\right]}\;\;,
\end{equation}  

\noindent where $\eta=\Omega_n/\omega$ and $\Omega_n = 2E_n/(\hbar n)$ is the Bohr
frequency of the electron in the circular orbit with Bohr radius $r_n=\hbar^2
n^2/(me^2)$.  The dipole approximation requires that $\omega r_n << c$, hence
$\eta>>(137n)^{-1}$; moreover, we note that $\hbar\omega>E_n$ implies that
$\eta<2/n$.  For $n=1$, the ground state is spherically symmetric and hence
$\sigma_-=\sigma_+$; however for $n>1$ the electron following the ``circular
orbit'' with $m=l=n-1>0$ tends to move on average like the observer in our thought
experiment and equation (17) implies that $\sigma_-<\sigma_+$, as expected.  

Further corroboration of the predictions of the nonlocal theory may be obtained
from the consideration of ``circular orbits'' of electrons in a uniform magnetic
field $B$ along the $z$-axis.  In fact, on the basis of Larmor's celebrated theorem
it would be natural to consider such electronic states in connection with our
thought experiment involving an observer in a rotating frame of reference. 
Ignoring any motion along the $z$-axis, such a ``circular orbit'' would be
characterized by the quantum numbers $N$ and $M$, where $N$ denotes the energy
states $\hbar\Omega_c(N+{1\over 2})$ and $\hbar M$ is the $z$-component of the
angular momentum of the electron.  Here $\Omega_c = eB/(mc)$ is the cyclotron
frequency and we assume that $\hbar\Omega_c<<mc^2$.  The correspondence with the
classical cyclotron motion of the electron with orbital frequency $\Omega_c$ can be
established for $N\sim M>>1$.  In this case, the nonrelativistic calculation of
electric dipole transitions due to a normally incident radiation of frequency
$\omega, \hbar\omega<< mc^2$, and definite helicity reveals that transition is
possible only for $(N, M)\rightarrow (N+1, M+1)$ with $\omega = \Omega_c$ and
positive helicity incident radiation, while for incident negative helicity
radiation of $\omega = \Omega_c$ the transition $(N, M)\rightarrow (N+1, M-1)$ is
forbidden.  This is in qualitative agreement with the result of the nonlocal
theory, equation (16), for the case of resonance $\omega = \Omega$ : At resonance,
the field amplitude for positive helicity radiation diverges linearly with proper
time, while the amplitude for negative helicity radiation remains finite.  Thus in
terms of the ratios of helicity amplitudes the result is qualitatively the same as
in quantum mechanics.

\section{Gravitation}  It appears from the results of the previous section that
the acceleration-induced nonlocality must be taken seriously, since its
predictions are closer to reality (as defined by the quantum theory) than the
standard theory of accelerated systems.  This circumstance raises the question of
whether the nonlocality extends to gravitation as would be intuitively expected
from Einstein's heuristic principle of equivalence.  

Gravitation is a universal interaction that is qualitatively different from other
interactions and so it may not be surprising if it could be described in terms of a
nonlocal classical field in Minkowski spacetime such that in a suitable eikonal
limit this nonlocal field would have an interpretation in terms of the local
curvature of a certain spacetime manifold as in general relativity.  

\section*{Acknowledgments} I thank C. Chicone for many valuable discussions
regarding the problem of uniqueness of the kinetic kernel.  I am grateful to I.
Ciufolini and L. Lusanna for their kind invitation and excellent hospitality.

\end{document}